\documentclass[12pt]{iopart}
\usepackage{psfig}
\usepackage{epsfig}

\newcommand{\be}{\begin{equation}}
\newcommand{\ee}{\end{equation}}
\newcommand{\bea}{\begin{eqnarray}}
\newcommand{\eea}{\end{eqnarray}}

\begin{document}
\title[Nonequilibrium Models of Relativistic Heavy-Ion Collisions]
{Nonequilibrium Models of Relativistic Heavy-Ion Collisions}

\author{H.~St\"ocker$^{1,2}$, E. L. Bratkovskaya$^1$,
M. Bleicher$^1$, S. Soff$^1$, and X. Zhu$^{1,2,3}$}

\address{$^1$ Institut f\"{u}r Theoretische Physik,
 Johann Wolfgang Goethe -- Universit\"{a}t,
 Robert Mayer Str. 8-10,
 60054 Frankfurt am Main, Germany}
\address{$^2$  Frankfurt Institute for Advanced Studies (FIAS),
 Robert Mayer Str. 8-10,
 60054 Frankfurt am Main, Germany}
\address{$^3$ Physics Department, Tsinghua University,
Beijing 100084, China}


\begin{abstract}
We review the results from the various hydrodynamical and transport
models on the collective flow observables from AGS to RHIC energies.
A critical discussion of the present status of the CERN experiments on
hadron collective flow is given. We emphasize the importance of the
flow excitation function from 1 to 50 A$\cdot$GeV:  here the
hydrodynamic model has predicted the collapse of the $v_1$-flow  and of
the $v_2$-flow at $\sim 10$ A$\cdot$GeV; at 40 A$\cdot$GeV it has been
recently observed by the NA49 collaboration.  Since hadronic
rescattering models predict much larger flow than observed at this
energy we interpret this observation as evidence for a first
order phase transition at high baryon density $\rho_B$.  Moreover, the
connection of the elliptic flow $v_2$ to jet suppression is examined.
It is proven experimentally that the collective flow is not faked by
minijet fragmentation. Additionally, detailed transport studies show
that the away-side jet suppression can only partially ($<$ 50\%) be due
to hadronic rescattering.
Furthermore, the change in sign of $v_1, v_2$ closer to beam
rapidity is related to the occurence of a high density first order
phase transition in the RHIC data at 62.5, 130 and 200 A$\cdot$GeV.
\end{abstract}



\section{Introduction: Old and new observables for the QGP phase transition}

Lattice QCD results \cite{Fodor04,Karsch04}
show a crossing, but no first order phase transition to the QGP for
vanishing or small chemical potentials $\mu_B$, i.e. at the conditions
accessible at central rapidities at RHIC full energies.  A first order
phase transition does occur according to the QCD lattice calculations
\cite{Fodor04,Karsch04} only at high baryochemical potentials or
densities, i.e.  at SIS-300 and lower SPS energies and in the
fragmentation region of RHIC, $y \approx 4-5$
\cite{Anishetty80,Date85}.  The critical baryochemical potential is
predicted \cite{Fodor04,Karsch04} to be $\mu_B^c \approx 400 \pm 50
\mbox{ MeV} $ and the critical temperature $T_c \approx 150 - 160$ MeV.
We do expect a phase transition also at finite strangeness.
Predictions for the phase diagram of strongly interacting matter for
realistic non-vanishing net strangeness are urgently needed to obtain a
comprehensive picture of the QCD phase structure.  Multi-Strangeness
degrees of freedom are very promising probes for the properties of the
dense and hot matter \cite{Koch86}. The strangeness distillation
process \cite{greic87,greic88} predicts dynamical de-admixture of $s$
and $\bar{s}$ quarks, which yields unique signatures for QGP creation:
high multistrange hyperon-/-matter production, strangelet formation and
unusual antibaryon to baryon ratios $ect$.

\begin{figure}[t]
\begin{minipage} [r] {6cm}
\psfig{file=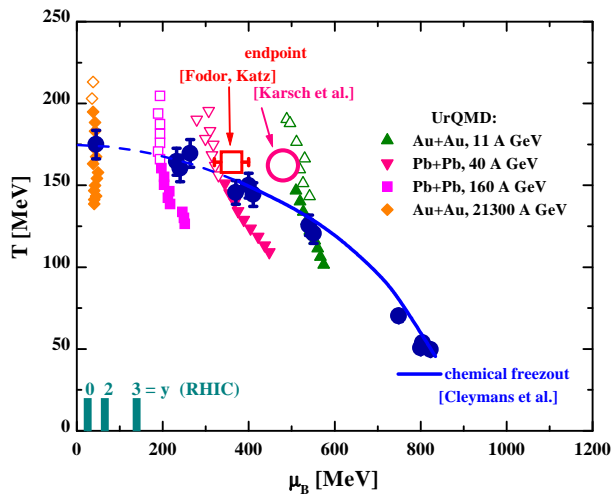,width=8cm}
\end{minipage}
\begin{minipage} [r] {9cm}
\caption{
The new phase diagram with the critical end point at $\mu_B \approx 400
\mbox{ MeV}, T \approx 160 \mbox{ MeV} $ as predicted by Lattice QCD
\protect\cite{Fodor04}.
In addition, the time evolution in the $T-\mu$-plane of a central cell
in UrQMD calculations \cite{Bravina} is depicted for
different bombarding energies.  Note, that the calculations indicate
that bombarding energies $E_{LAB} \stackrel{<}{\sim} 40$ A$\cdot$GeV are
needed to probe a first order phase transition. At RHIC (see insert at
the $\mu_B$ scale) this point is accessible in the fragmentation region
only (taken from \protect{\cite{Bratkov04}}).} \label{phasedia}
\end{minipage}
\end{figure}

A comparison of the thermodynamic parameters $T$ and $\mu_B$ extracted
from the UrQMD-transport model in the central overlap regime of Au+Au
collisions \cite{Bratkov04} with the QCD predictions is shown in Fig 1,
where the full dots with errorbars denote the 'experimental' chemical
freeze-out parameters -- determined from fits to the experimental
yields -- taken from Ref.  \cite{Cleymans}. The triangular and
quadratic symbols (time-ordered in vertical sequence) stand for
temperatures $T$ and chemical potentials $\mu_B$ extracted from UrQMD
transport calculations in central Au+Au (Pb+Pb) collisions at RHIC
(21.3 A$\cdot$TeV), 160, 40 and 11 A$\cdot$GeV \cite{Bravina} as a
function of the reaction time (separated by 1 fm/c steps from top to
bottom).  The open symbols denote nonequilibrium configurations and
correspond to $T$ parameters extracted from the transverse momentum
distributions, whereas the full symbols denote configurations in
approximate pressure equilibrium in longitudinal and transverse
direction.

During the nonequilibrium phase (open symbols) the transport
calculations show much higher temperatures (or energy densities) than
the 'experimental' chemical freeze-out configurations at all bombarding
energies ($\geq$ 11 A$\cdot$GeV).
These numbers are also higher than the critical point (circle) of (2+1)
flavor - Lattice QCD calculations by the
Bielefeld-Swansea-collaboration  \cite{Karsch04} (large open circle)
and by the Wuppertal-Budapest-collaboration \cite{Fodor04}.
The energy density at $\mu_c, T_c$ is in the order of  $\approx$ 1
GeV/fm$^3$ (or slightly below).  At RHIC energies a cross-over is
expected at midrapidity, when stepping down in temperature during the
expansion phase of the 'hot fireball'. The baryon chemical potential
$\mu_B$ for different rapidity intervals at RHIC energies has been
obtained from a statistical model analysis by the BRAHMS Collaboration
based on measured antihadron to hadron yield ratios
\cite{BRAHMS_PRL03}.  For midrapidity one finds $\mu_B\simeq 0$,
whereas for forward rapidities $\mu_B$ increases up to $\mu_B\simeq
130$~MeV at $y=3$.  Thus, only extended forward rapidity measurement
($y \approx 4-5)$ will allow to probe large $\mu_B$ at RHIC.  The
detectors at RHIC at present offer only a limited chemical potential
range.  This situation changes at lower SPS (and top AGS) as well as at
the future GSI SIS-300 energies: sufficiently large chemical potentials
$\mu_B$ should allow for a first order phase transition \cite{Shuryak}
(to the right of the critical point in the ($T, \mu_B$) plane).  The
transport calculations show high temperatures (high energy densities)
in the very early phase of the collisions, only. Here, hadronic
interactions are weak due to formation time effects and yield little
pressure.  Diquark, quark and gluon interactions should cure this
problem.

\section{Directed and elliptic flow}

\subsection{General considiration}

Hydrodynamic flow and shock formation has been proposed early
\cite{Hofmann74,Hofmann76} as the key mechanism for the creation of
hot and dense matter during relativistic heavy-ion collisions.  The
full three-dimensional hydrodynamical flow problem is much more
complicated than the one-dimensional Landau model \cite{Landau}:  the
3-dimensional compression and expansion dynamics yields complex triple
differential cross-sections, which provide quite accurate spectroscopic
handles on the equation of state.  The bounce-off, the
squeeze-out  and the antiflow
\cite{Stocker79,Stocker80,Stocker81,Stocker82,Stocker86} (third flow
component \cite{Csernai99,Csernai04}) serve as differential barometers
for the properties of compressed, dense matter from SIS to RHIC.
Presently, the most employed flow observables are \cite{Voloshin}:
\bea
v_1 = \left< \frac{p_x}{p_T} \right>, \ \ \ v_2 = \left<
\frac{p_x^2 - p_y^2}{p_x^2 + p_y^2} \right> \, .
\eea
Here, $p_x$ denotes the momentum in
$x$-direction, i.e. the transversal momentum within the reaction plane
and $p_y$ the transversal momentum out of the reaction plane. The total
transverse momentum is given as $p_T = \sqrt{p_x^2 + p_y^2}$; the
$z$-axis is in the beam direction.  Thus,  $v_1$ measures the
''bounce-off'', i.e. the strength of the directed flow in the reaction
plane, and $v_2$ gives the strength of the second moment of the
azimuthal particle emission distribution, i.e.
''squeeze-out'' for  $v_2 < 0$
\cite{Hofmann74,Hofmann76,Stocker79,Stocker80,Stocker81,Stocker82,Stocker86}.
In particular,  it has been shown
\cite{Hofmann76,Stocker79,Stocker80,Stocker81,Stocker82,Stocker86} that
the disappearence or ''collapse'' of flow is a direct result of a first
order phase transition.

Several hydrodynamic models have been used in the past, starting with
the one-fluid ideal hydrodynamic approach.  It is well known that the
latter model predicts far too large flow effects. To obtain a better
description of the dynamics, viscous fluid models have been developed
\cite{Schmidt93,Muronga01,Muronga03}.  In parallel, so-called
three-fluid models, which distinguish between projectile, target and
the fireball fluids, have been considered \cite{Brachmann97,Toneev03}.
Here viscosity effects appear only between the different fluids, but
not inside the individual fluids. The aim is to have at our disposal a
reliable, three-dimensional, relativistic three-fluid model including
viscosity \cite{Muronga01,Muronga03}.

Flow can be described very elegantly in hydrodynamics
(cf. Refs. \cite{Kolb,Teaney,Shuryak1,Nara}) by a proper choice of initial
conditions which have very strong influence on the final results.
In this respect, it is important to consider also microscopic multicomponent
(pre-) hadron transport theory, e.g.  models like qMD \cite{Hofmann99},
IQMD \cite{Hartnack89}, RQMD \cite{Sorge}, UrQMD \cite{Bass98} or HSD
\cite{Cassing99}, as control models for viscous hydro and as background
models to subtract interesting non-hadronic effects from data.  If
Hydro with and without quark matter EoS, hadronic transport models
without quark matter -- but with strings -- are compared to data, can
we learn whether quark matter has been formed?  What degree of
equilibration has been reached? What does the equation of state look
like?  How are the particle properties, self energies, cross sections
changed?

To estimate systematic model uncertainties, the results of the
different microscopic transport models also have to be carefully
compared. The two robust hadron/string based models, HSD and UrQMD, are
considered in the following.

\subsection{Review of AGS and SPS results}

Microscopic (pre-)hadronic transport models describe the formation and
distributions of many hadronic particles at AGS and SPS rather well
\cite{Weber02}.  Furthermore, the nuclear equation of state has been
extracted by comparing to flow data which are described reasonably well
up to AGS energies
\cite{Andronic03,Andronic01,Soff99,Csernai99,Sahu1,Sahu2}.  Ideal hydro
calculations, on the other hand, predict far too much flow at these
energies \cite{Schmidt93}.  Thus, viscosity effects have to be taken
into account in hydrodynamics.

In particular, ideal hydro calculations are factors of two higher than
the measured sideward flow at SIS \cite{Schmidt93} and AGS, while the
directed flow $p_x/m$ measurement of the E895 collaboration shows that
the $p$ and $\Lambda$ data are reproduced reasonably well \cite{Soff99}
in UrQMD, i.e. in a hadronic transport theory with reasonable
cross-sections, i.e.  realistic mean-free-path of the constituents.

Only ideal hydro calculations predict, however, the appearance of a
so-called ''third flow component'' \cite{Csernai99} or ''antiflow''
\cite{Brach00} in central collisions. We stress that this only holds if
the matter undergoes a {\it first order phase transition} to the QGP. The
signal is that around midrapidity the directed flow, $p_x (y)$, of
protons develops a negative slope! In contrast, a hadronic EoS without
QGP phase transition does not yield such an exotic ''antiflow''
(negative slope) wiggle in the proton flow $v_1(y)$.

\begin{figure}[t]
\begin{minipage}[l]{6 cm}
\psfig{file=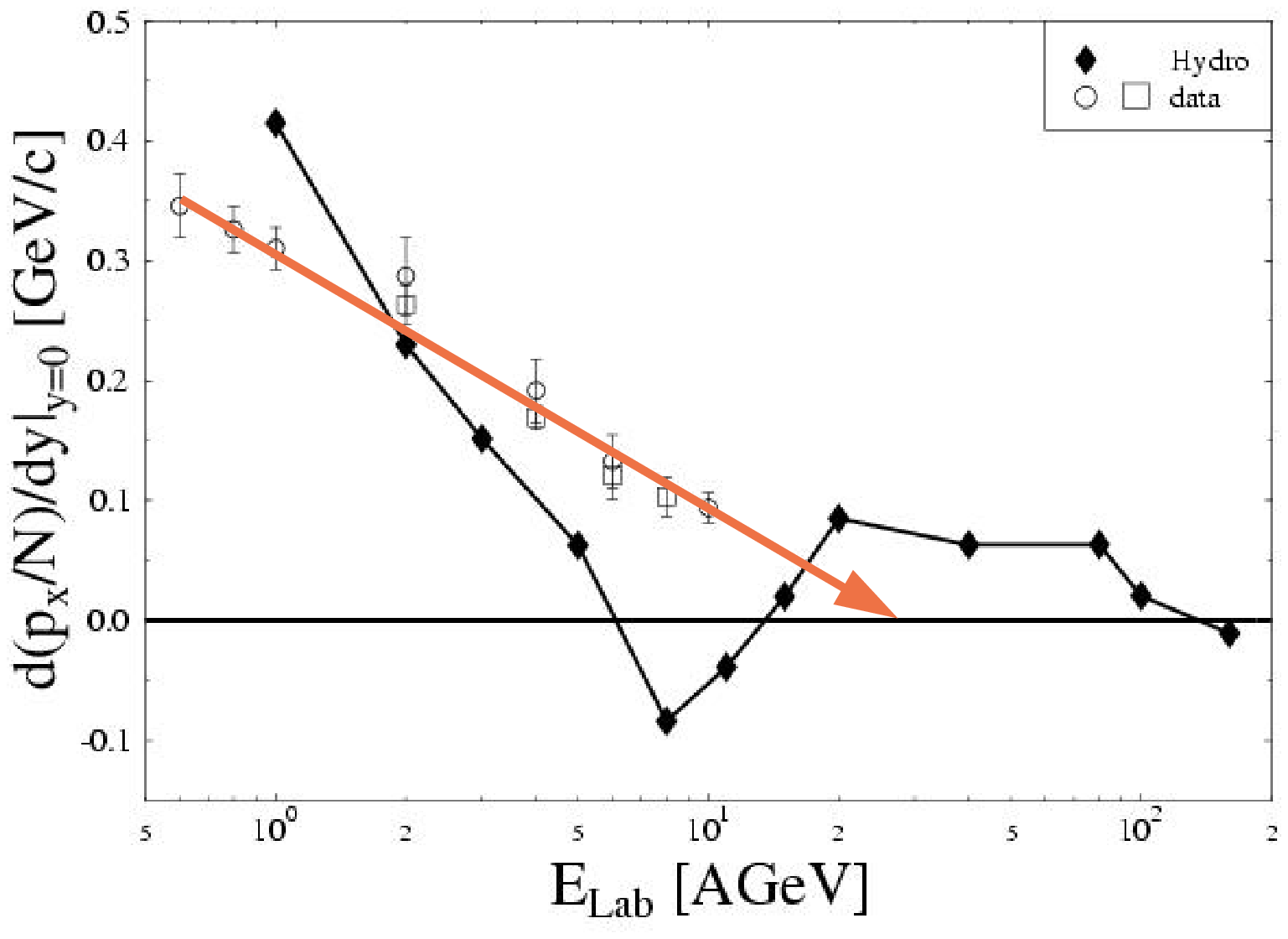,width=7.5cm,clip}
\end{minipage}
\begin{minipage}[r]{10 cm}
\hspace*{2cm}
\psfig{file=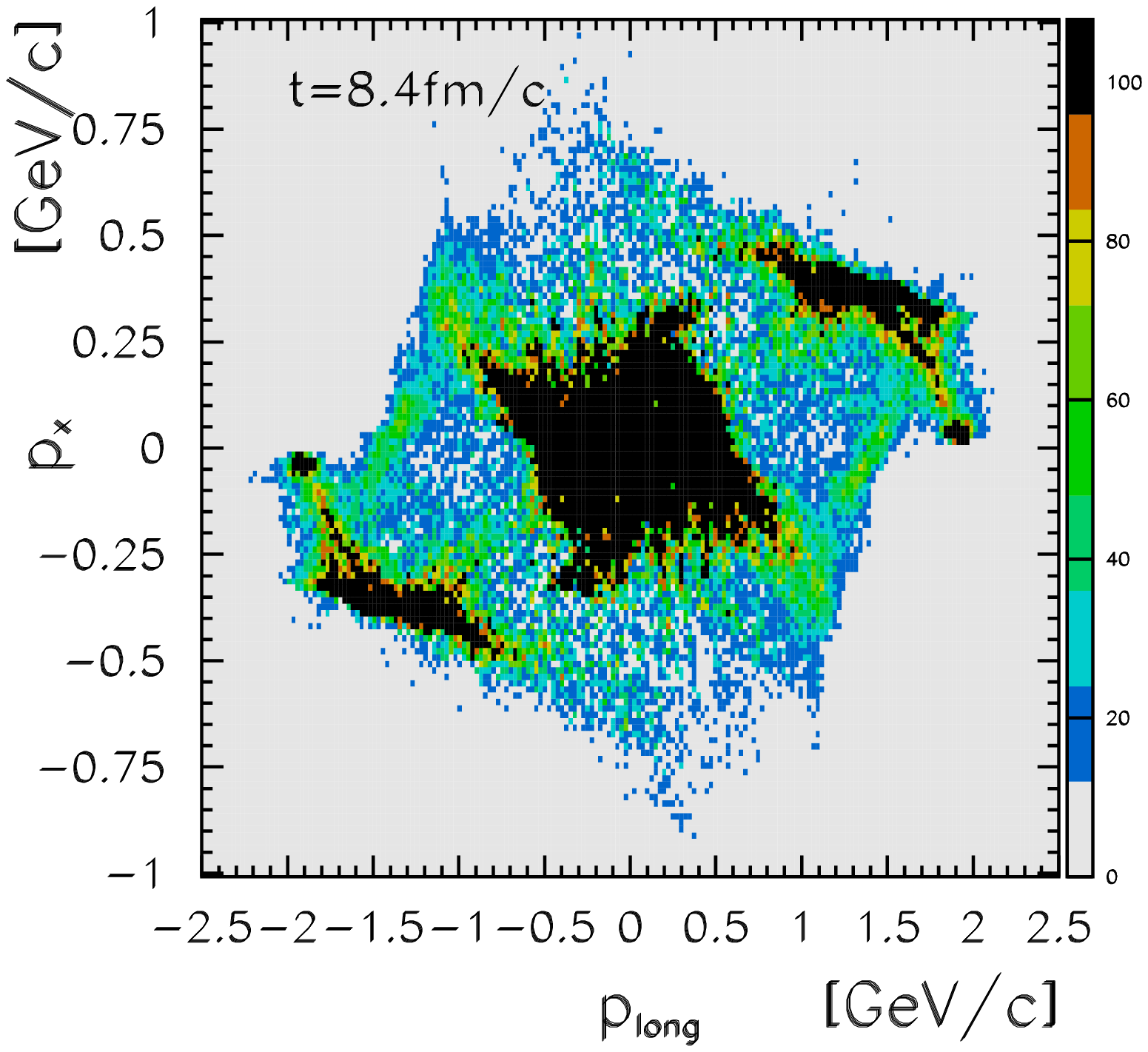,scale=0.4,clip}
\end{minipage}
\caption{Left: Measured SIS and AGS proton $dp_x/dy$-slope data
compared to a three-fluid hydro calculation \protect{\cite{Paech00}.  A
linear extrapolation of the AGS data indicates a collapse of flow at
$E_{Lab} \approx 30$ A$\cdot$GeV, i.e. for the lowest SPS- and the
upper FAIR- energies at GSI.
Right: Net-baryon density in momentum space for Pb+Pb at 8 A$\cdot$GeV
for b=3 fm at time 8.4 fm/c calculated in three-fluid hydro
\cite{Paech00} for condition $s/\rho < 10$.}
\label{flow_extra}}
\end{figure}

The ideal hydrodynamic directed proton flow $p_x$  (Fig.
\ref{flow_extra}) shows even negative values between 8 and 20
A$\cdot$GeV. An increase back to positive flow is predicted with
increasing energy, when the compressed QGP phase is probed.  But, where
is the predicted minimum of the proton flow in the data?  The hydro
calculations suggest that this ''softest point collapse'' is at $E_{Lab}
\approx 8$ A$\cdot$GeV.  This has not been verified by the AGS data!
However, a linear extrapolation of the AGS data indicates a collapse of
the directed proton flow at $E_{Lab} \approx 30$ A$\cdot$GeV (Fig.
\ref{flow_extra}).

Recently, substantial support for this prediction has been obtained by
the low energy 40 A$\cdot$GeV SPS data of the NA49 collaboration
\cite{NA49_v2pr40} (Fig. \ref{v1v2_40}).  These data clearly
show the first proton ''antiflow'' around mid-rapidity, in contrast to
the AGS data as well as to the UrQMD and HSD calculations involving no
phase transition (Fig. \ref{v1v2_40}, l.h.s.).  Thus, at bombarding
energies of 30-40 A$\cdot$GeV, a first order phase transition to the
baryon rich QGP most likely is already observed; the first order phase
transition line is crossed (cf. Fig.  \ref{phasedia}). This is the
energy region where the new FAIR- facility at GSI will operate.  There
are good prospects that the baryon flow collapses and other first order
QGP phase transition signals can be studied at the lowest SPS energies
as well as at the RHIC fragmentation region $y > 4-5$.  These
experiments will enable a detailed study of the first order phase
transition at high $\mu_B$ and of the properties of the baryon rich
QGP.

\section{Proton elliptic flow collapse at 40 A$\cdot$GeV - evidence for a
first order phase transition at highest net baryon densities}

\begin{figure}[t]
\vspace*{5mm}
\centerline{\psfig{file=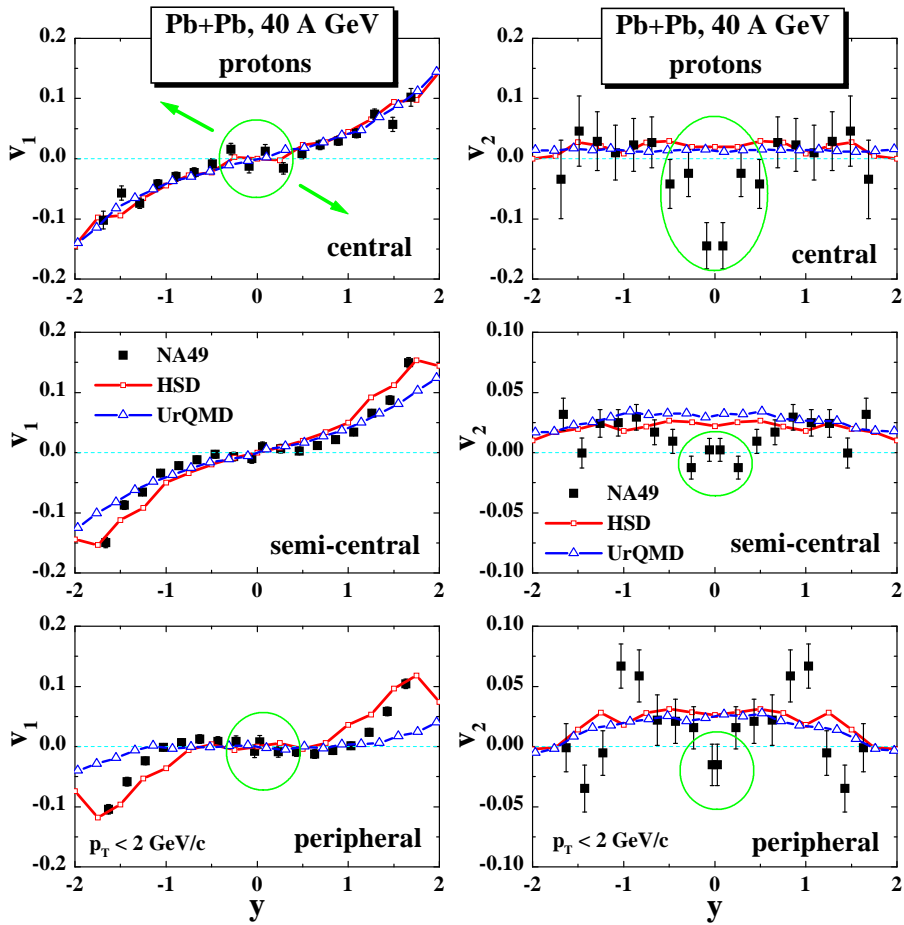,scale=0.62}}
\caption{Proton directed $v_1$ (left) and elliptic $v_2$ (right) flow
for central, semi-central and peripheral Pb+Pb collissions at 40
A$\cdot$GeV. The full squares indicate NA49 data
\protect\cite{NA49_v2pr40}, the solid lines with open squares show the
HSD results whereas the solid lines with open triangles are the
UrQMD results.}
\label{v1v2_40}
\end{figure}

At SIS energies microscopic transport models reproduce the data on the
excitation function of the proton elliptic flow $v_2$ quite well:  A
soft, momentum-dependent equation of state
\cite{Andronic00,Andronic99,Larionov} seems to account for the data.
The observed proton flow $v_2$ below $\sim$ 5 A$\cdot$GeV is smaller than
zero, which corresponds to the squeeze-out predicted by hydrodynamics
long ago
\cite{Hofmann74,Hofmann76,Stocker79,Stocker80,Stocker81,Stocker82,Stocker86}.
The AGS data exhibit a transition from
squeeze-out to in-plane flow in the midrapidity region.  The change in
sign of the proton $v_2$ at 4-5 A$\cdot$GeV is in accord with transport
calculations -- UrQMD \cite{Soff99} and HSD \cite{Sahu1,Sahu2}).  At
higher energies, 10-160 A$\cdot$GeV, a smooth increase of the flow
$v_2$ is predicted from the string-hadronic transport models.  In fact,
the 158 A$\cdot$GeV data of the NA49 Collaboration suggest that this
smooth increase proceeds between AGS and SPS as predicted.

This is in strong contrast to recent NA49 data at 40 A$\cdot$GeV (cf.
Fig. \ref{v1v2_40}, r.h.s.): A sudden collapse of the proton flow $v_2$
is observed for central, midcentral as well as for peripheral protons.
This collapse of $v_2$ for protons around midrapidity at 40 A$\cdot$GeV
is very pronounced while it is not observed at 158 A$\cdot$GeV. The
UrQMD and HSD calculations, without a phase transition, show a robust,
but wrong ~3\% flow of protons - in strong contrast to the data.

Thus,  the collapse of the $v_1$ and $v_2$ flow has been observed by
NA49 \cite{NA49_v2pr40} at the same energy around 40 A$\cdot$GeV.  This
is the highest energy -- according to \cite{Fodor04,Karsch04} and Fig.
\ref{phasedia} -- at which a first order phase transition can be
reached at the central rapidities of relativistic heavy-ion collisions.
We, therefore, conclude that a first order phase transition at the
highest baryon densities accessible in nature has been seen at these
energies in Pb+Pb collisions.

\subsection{Strong collective flow at RHIC signals a new phase of matter}

The rapid thermalization obtained in parton cascade calculations by Xu
and Greiner \cite{Xu04} by including three-body processes $gg
\leftrightarrow ggg $ in leading-order pQCD (besides gluon- and quark-
two-body elementary parton-parton scatterings) justifies {\it a
posteriori} the use of hydrodynamical calculations for the time
evolution of the complex four-dimensional expansion of the plasma.
However, there is no justification for the use of simple ideal
hydrodynamics (i.e.  neglecting the important transport coefficients)
and simple, smooth initial conditions in hydrodynamics
\cite{Muronga01,Muronga03,Teaney1}.  PHOBOS data at $\sqrt{s}$= 130 GeV
and 200 GeV suggest energy independent $v_2 (\eta)$ distributions.
Furthermore, the observed distribution has a triangular shape in
rapidity.  This experimental finding is in strong disagreement with
Bjorken boost invariant hydro predictions \cite{Heinz04,Shuryak}, which
fit only the midrapidity region.

The predicted average proton $v_2$-values obtained from the SPHERIO
hydro code  with NEXUS initial conditons \cite{Aguiar01}) are by
factors of two higher than simple smooth initial state hydrodynamic
calculations.  This indicates that ideal hydro with naive smooth
initial conditions -- as used by many authors -- do not describe but
rather fit the data.  Strong viscosity effects can play a role for
particles with $p_T < 1.2$ GeV/c: a decent description of the dynamics
requires, however, relativistic viscous hydro simulations
\cite{Muronga01,Muronga03,Muronga03b}.  The NexSpherio simulations
\cite{Aguiar01} predict very large event-by-event fluctuations of $v_2$
caused by the strongly fluctuating initial conditions (given by NEXUS).
This effect has been also studied in Ref. \cite{Snelling} where the
authors found a strong influence of spatial eccentricity fluctuations
on the determination of elliptic flow.

\begin{figure}[!]
\begin{minipage}[l]{7 cm}
\psfig{file=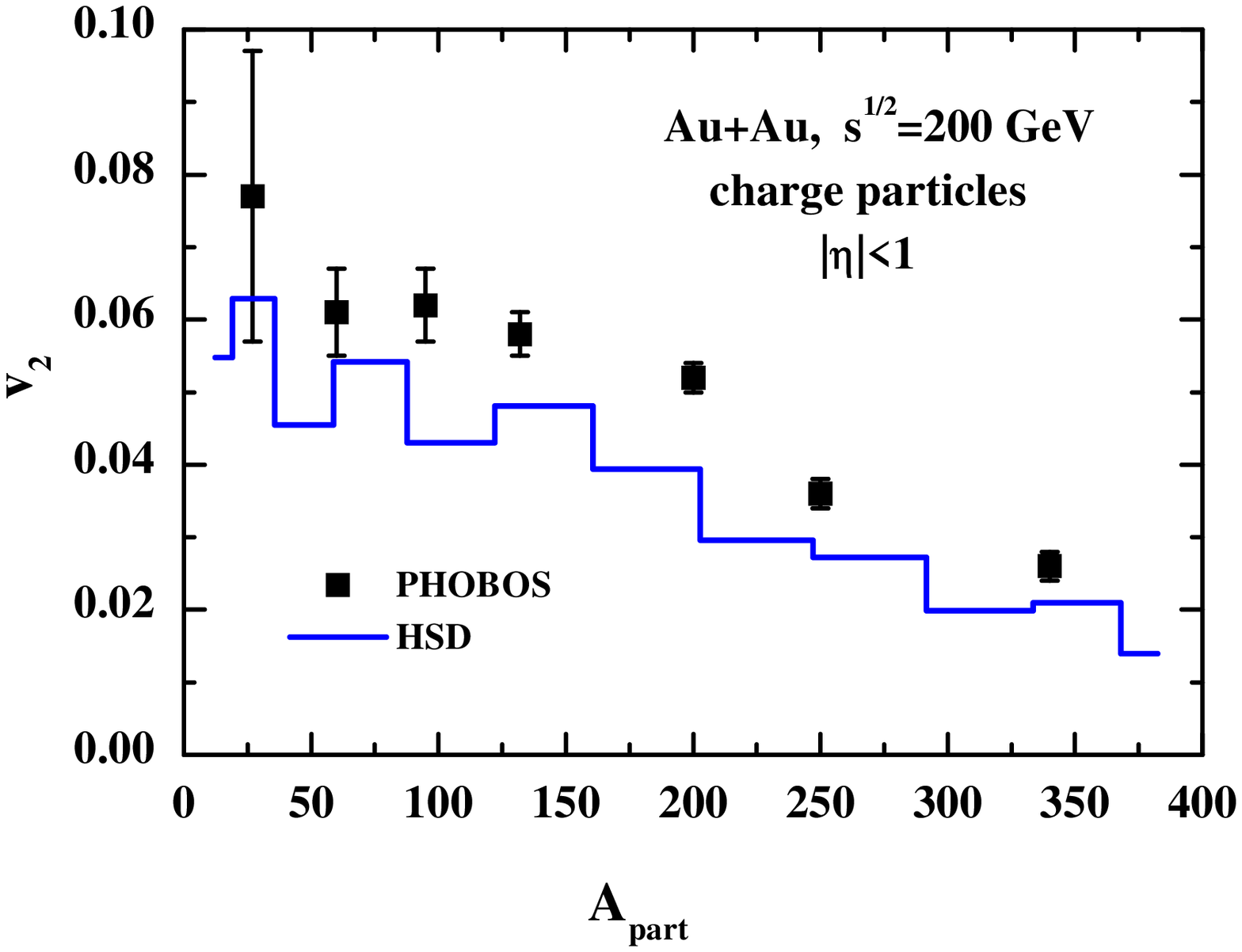,scale=0.4,clip}
\end{minipage}
\begin{minipage}[r]{8 cm}
\hspace*{0.5cm}
\psfig{file=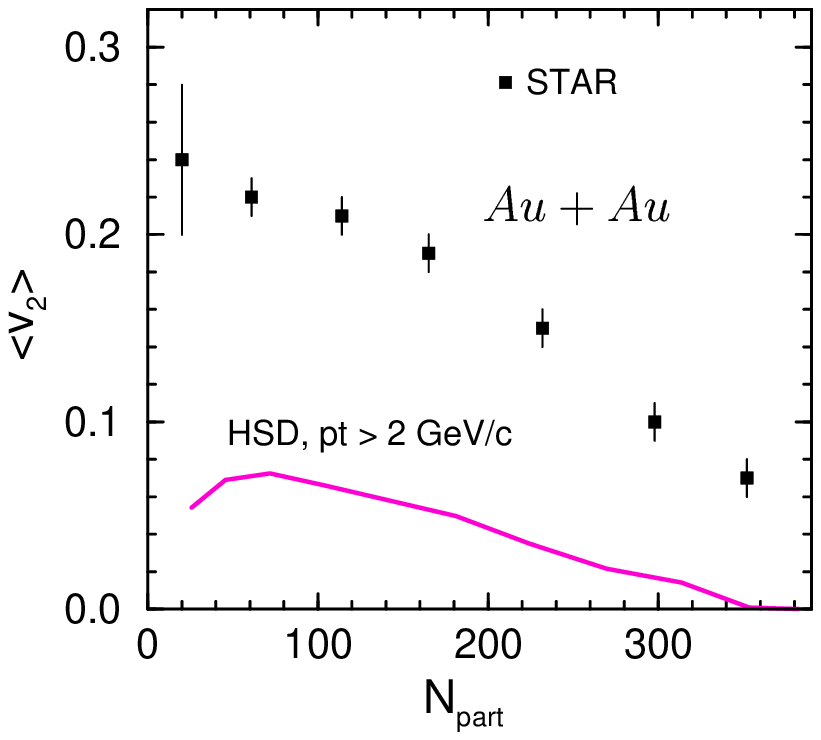,scale=0.7,clip}
\end{minipage}
\caption{
Left: The HSD result \cite{Brat03} for the elliptic flow $v_2$ for
charged hadrons as a function of the number of 'participating nucleons'
for $|\eta| \leq$ 1 for $Au+Au$ collisions at $\sqrt{s}$ = 200 GeV
in comparison to the 'hit-based analysis' data of the
PHOBOS Collaboration \protect\cite{PHOBOS1}.
Right: The HSD result \protect{\cite{CGG}} for $v_2$ for charged hadrons
with $p_T > 2$ GeV/c as a function of the number of 'participarting
nucleons' in comparison to the STAR data.}
\label{v2_rhic}
\end{figure}

Microscopic transport simulations (HSD and UrQMD) of particle yields,
$dN/dy$ distributions, etc.  give a reasonable description of the RHIC Au+Au
data \cite{Bratkov04,Brat03}. The HSD and UrQMD transport approaches
are based on string, quark, diquark ($q, \bar{q}, qq, \bar{q}\bar{q}$)
as well as hadronic degrees of freedom but lack explicit gluonic
degrees of freedom. At RHIC, UrQMD and HSD yield reasonable abundances
of light hadrons composed of $u,d,s$ quarks \footnote{For a more recent
survey on hadron rapidity distributions from 2 to 160 A$\cdot$GeV in
central nucleus-nucleus collisions within the HSD and UrQMD transport
approaches we refer the reader to Ref.  \cite{Weber02}.}.
Do they also predict the collective flow properly?

\begin{figure}[t]
\begin{minipage}[l]{7 cm}
\psfig{file=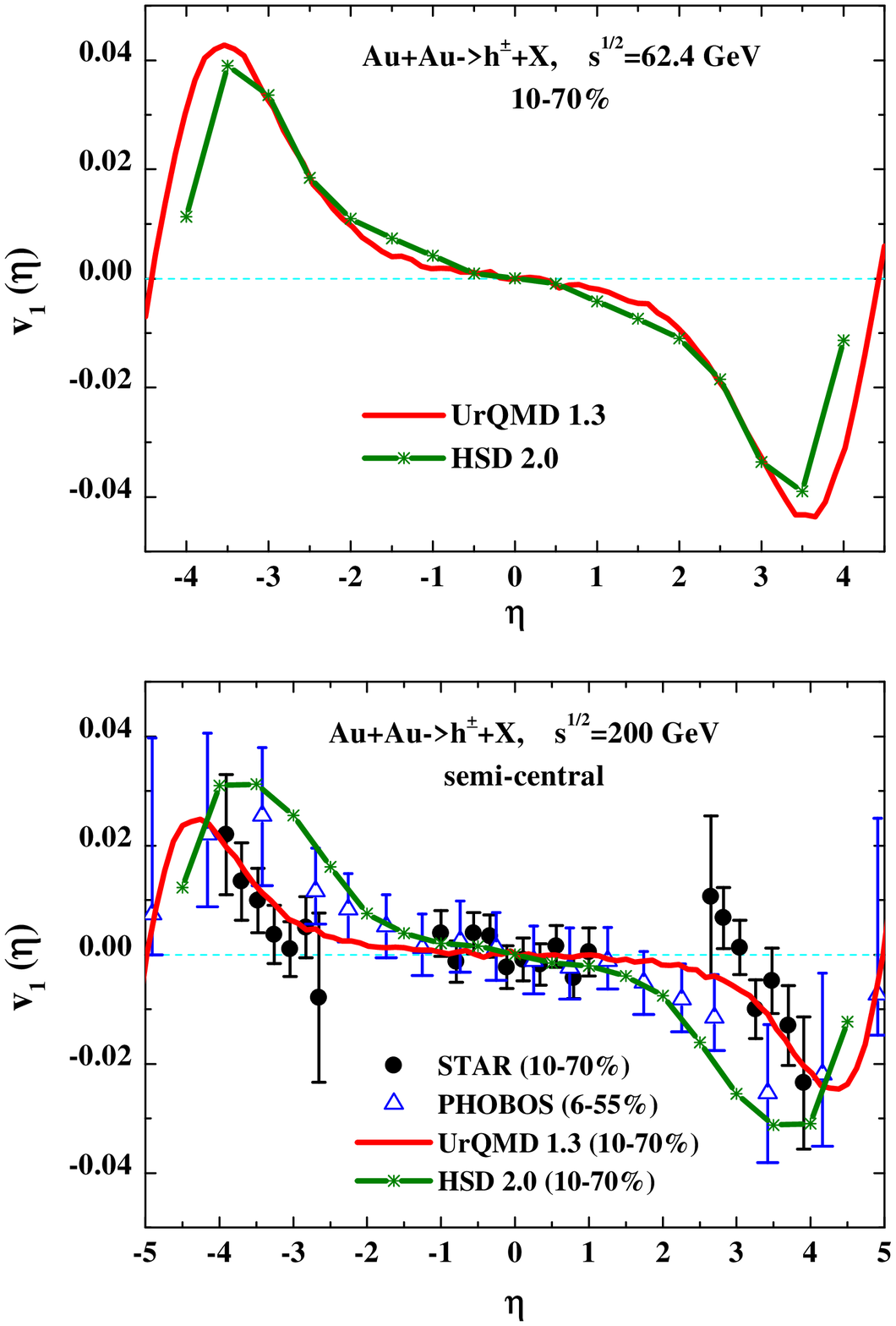,scale=0.5,clip}
\end{minipage}
\begin{minipage}[r]{8 cm}
\phantom{a}\vspace*{-5mm}\hspace*{8mm}
\psfig{file=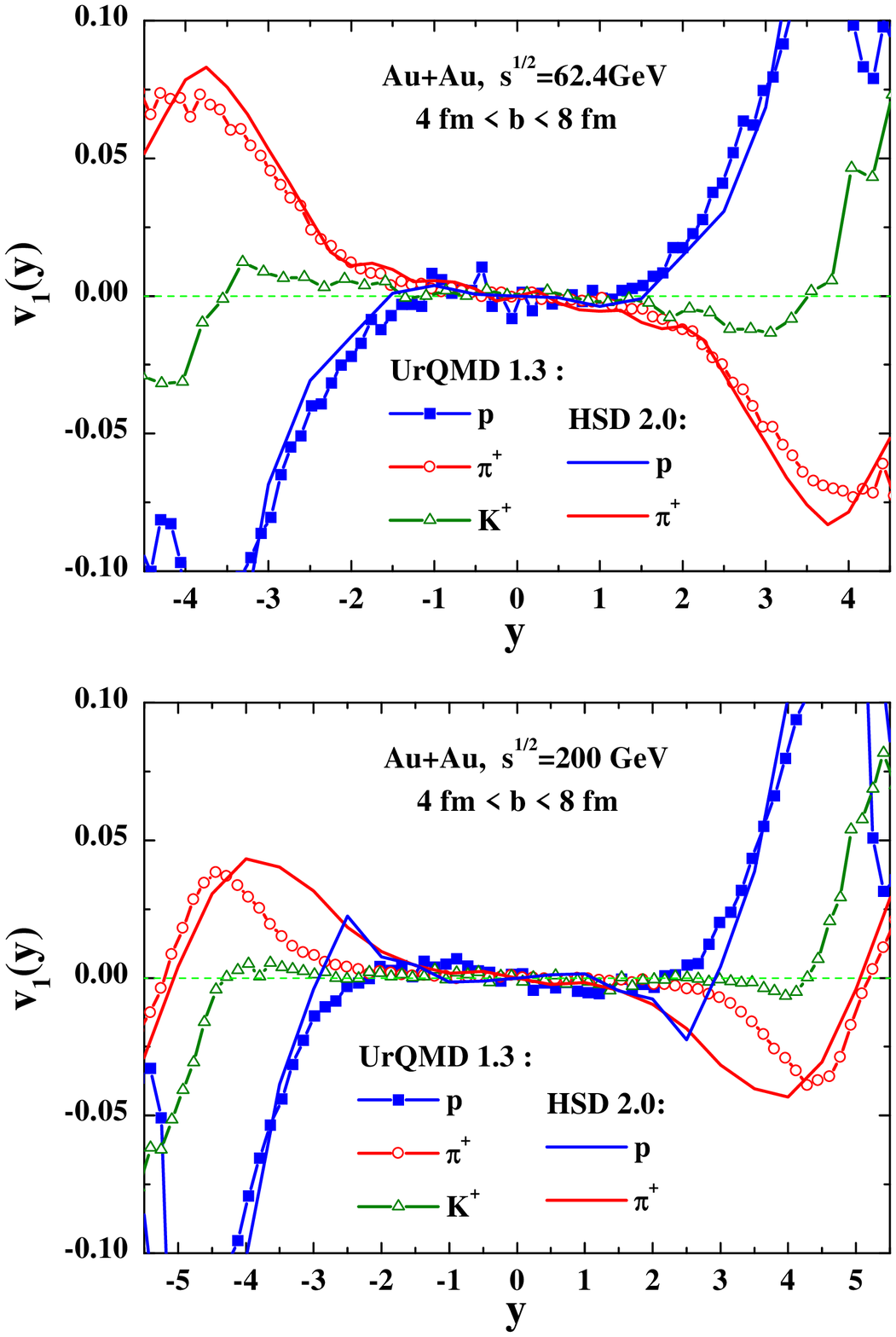,scale=0.5,clip}
\end{minipage}
\caption{Left: The directed flow $v_1$ for  charged hadrons from
 semi-central Au + Au collisions at $\sqrt{s}=$62.4 GeV (upper plot)
and 200 GeV (lower plot) versus pseudorapidity  $\eta$ in comparison
to the data from STAR \cite{STAR_v1Heta} (10-70\% centrality, solid dots)
and PHOBOS \cite{PHOBOS_v1Heta} (6-55\% centrality, open triangles) at
$\sqrt{s}=200$ GeV.  The solid lines and lines with stars correspond to
the UrQMD 1.3 and HSD 2.0 results.  Right:  UrQMD 1.3 (lines with
symbols)and HSD 2.0 (solid lines) results for the $v_1$ for protons,
$\pi^+$ and $K^+$ from semi-central Au + Au collisions at
$\sqrt{s}=$62.4 GeV (upper plot) and 200 GeV (lower plot) versus
rapidity $y$.}
\label{Fig_Zhu}
\end{figure}

The left part of Fig. \ref{Fig_Zhu} shows the UrQMD 1.3 and HSD 2.0
results for the directed flow $v_1$ for charged hadrons from
semi-central Au + Au collisions at $\sqrt{s}=$200 GeV (left lower plot)
versus pseudorapidity  $\eta$ in comparison to the data from STAR
\cite{STAR_v1Heta} (10-70\% centrality, solid dots) and PHOBOS
\cite{PHOBOS_v1Heta} (6-55\% centrality, open triangles) at
$\sqrt{s}=200$ GeV.  The  upper left plot in Fig. \ref{Fig_Zhu}
presents the UrQMD and HSD predictions for $\sqrt{s}=$62.4 GeV.
UrQMD 1.3 gives a lower $v_1$ as compared to HSD due to the missing
jet production in this version. For the UrQMD 2.0 results (which
include PYTHIA similar to HSD) we refer to Ref. \cite{BleicherSQM04}.
The right part of Fig. \ref{Fig_Zhu} present the
UrQMD 1.3 and HSD 2.0 results for $v_1(y)$ for protons, $\pi^+$ and
$K^+$ from semi-central Au + Au collisions at $\sqrt{s}=$62.4 GeV
(upper plot) and 200 GeV (lower plot). This shows that the charged
particle flow (left part of Fig. \ref{Fig_Zhu}) can be dominantly
attributed to pions.
The proton $v_1$ is closer to zero in UrQMD 1.3, while it shows
a small "antiflow" in HSD 2.0.
Further high statistics RHIC data will clarify the situation
with the directed flow from the experimental side.

The UrQMD prediction for the elliptic flow is clearly not compatible
with the measured ~6\% $v_2$ - it is sizeably underestimated
\cite{Bleicher00}.  When shortening the formation time
\cite{Bleicher00} one can  get the model results closer to the data,
but more additional initial pressure --
needed to create the missing extra flow -- is not justified in  the
hadronic transport models.

The eliptic flow $v_2$ at low transverse momenta (Fig. \ref{v2_rhic}
l.h.s.) is underestimated in the HSD model by $\sim$ 30\% \cite{Brat03}.
However, at high transverse momenta ($p_T > 2$ GeV/c) the
$v_2$-flow is underestimated even by a factor of three (Fig.
\ref{v2_rhic}, r.h.s.) in the HSD model \cite{CGG}.
The HSD results are very similar to those of the
hadronic rescattering model by Humanic et al.  \cite{Tom1,Tom2}
and agree with the calculations by Sahu et al.  \cite{Sahu02} performed
within the hadron-string cascade model JAM \cite{JAM}.
We mention that the microscopic quark-gluon-string model
\cite{Zabrodin04} inserts in addition short distance vector repulsion
in order to achieve high flow values.
Thus, the "missing" elliptic flow (as well as the inverse slopes
\cite{Bratkov04}) in hadron-string based models indicate that
effective partonic degrees of freedom in the initial phase are needed
to supply the large pressure and early strong interaction rate.

\section{High $p_T$ suppression}

\subsection{How much quenching of high $p_T$ hadrons is due to (pre-)
hadronic final state interactions?}

A (mini-)jet at RHIC can produce hard particles, with $p_T$ above 5
GeV/c, but must also form soft particles with $p_T$ around 2 GeV/c.  Jets
produced in the center of the plasma zone have to pass first through
the parton phase at very high temperatures, then through the correlated
diquark and constituent quarks and finally through the  hadronic phase
that has build up preferentially close to the surface of the fireball.
Very high $p_T$ jets with $\gamma > 10$ materialize only far outside of
the plasma.  Most of the jets -- observed at RHIC -- are at $p_T
\approx 4-5$~GeV/c.  More than 50\% of the leading jet particles at $p_T
\sim 5\,$GeV/c are baryons.  Pion jets of 5 GeV have a $\gamma \simeq 35$,
i.e., they form far outside the plasma.  However,
HSD-PYTHIA-calculations \cite{Cassing04} show that many pions stem from
decaying rho-jets.  But, $\rho$'s and protons of 5 GeV have $\gamma \simeq
5$. Thus, $\rho$ and p-jets hadronize with roughly 50\% probability
\cite{CGG,Kai} while passing through the expanding bulk matter.
We point out that all partonic and hadronic models have failed by
factors of 5-10 to predict the observed high baryon abundance.

The PHENIX \cite{PHENIX1} and STAR \cite{STAR1} collaborations reported
a suppression of meson spectra for transverse momenta $p_T$ above $\sim
3\,$GeV/c.  This suppression is not observed in d+Au interactions at the
same bombarding energy per nucleon \cite{E1,E2} and presents clear
evidence for the presence of a new form of matter.  However, it is not
clear at present how much of the observed suppression can be attributed
to (pre-)hadronic interactions (FSI) \cite{CGG,Kai}.  (In-)elastic
collisions of (pre-)hadronic high momentum states with some of the bulk
(pre-)hadrons in the fireball can contribute in particular to the
attenuation of $p_T \approx 5\,$GeV/c  transverse momentum hadrons at
RHIC \cite{Cassing04}:  Most of the medium momentum (pre-)hadrons from
a $\pm 5$ GeV/c double jet will materialize inside the dense plasma;
their transverse momenta being 0-4 GeV/c.  The particles are dominantly
$\rho$'s, K's and baryons at $p_T >2.5\,$GeV/c -- hence their formation
time is $\gamma \tau_F \approx 4$ fm/c in the plasma rest frame.  The
time for color neutralization can also be very small \cite{Kopel4} for
the leading particle due to early gluon emission.

The (pre-)hadronic interactions with the bulk of the (pre-)hadronic
comovers then must have clearly an effect: they, too, suppress the
$p_T$-spectrum \cite{CGG}.  (In)elastic reactions of the fragmented
(pre-)hadrons with (pre-)hadrons of the bulk system cannot be described
by pQCD:  The relevant energy scale $\sqrt{s}$ is a few GeV.  Such
(in-)elastic collisions are very efficient for energy degradation since
many hadrons with lower energies are produced.  On the average, 1 to 2
such interactions can account for up to 50\% of the attenuation of high
$p_T$ hadrons at RHIC \cite{Kai}.  Hence, the  hadronic fraction of the
jet-attenuation  had to be addressed.

Such studies have been carried out in Ref. \cite{CGG} within
the HSD transport approach \cite{Cassing99}.
Moderate to high transverse momenta ($>1.5$ GeV/c) have been
incorporated by a superposition of $p+p$ collisions described via
PYTHIA \cite{PYTHIA}.  We point out that in Au+Au collisions the
formation of secondary hadrons is not only controlled by the formation
time $\tau_f$, but also by the energy density in the local rest frame,
i.e. hadrons are not allowed to be formed if the energy
density is above 1 GeV/fm$^3$\footnote{This energy density cut is
employed in the default HSD approach.}.  The interactions of the leading
and energetic (pre-)hadrons with the soft hadronic and bulk matter are
thus explicitly modeled.

\begin{figure}[t]
\begin{minipage}[l]{7 cm}
\centerline{\psfig{file=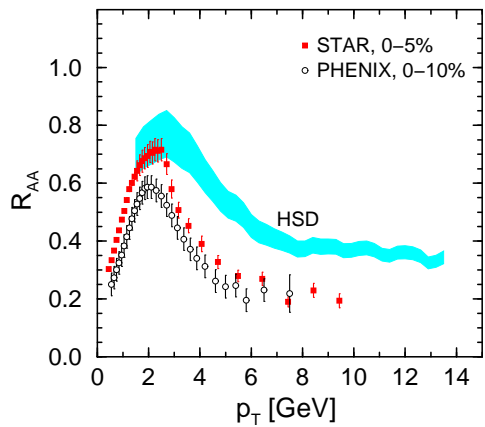,scale=0.35,angle=-90}}
\end{minipage}
\hspace*{-5mm}
\begin{minipage}[r]{9 cm}
\caption{The suppression factor $R_{AA}$ (3) of charged hadrons
at 5\% (10\%) central $Au+Au$ collisions ($\sqrt{s}$=200 GeV)
at midrapidity  (hatched band).
The experimental data are from Refs. \protect\cite{PHENIX2,STAR} and show
clearly that an additional  partonic suppression is needed
(taken from Ref. \protect\cite{CGG}).}
\end{minipage}
\label{fig5}
\end{figure}

Fig. \ref{fig5} shows the nuclear modification
factor \cite{Cassing04}
\begin{equation}
  \label{ratioAA}
  R_{\rm AA}(p_T) = \frac{1/N_{\rm AA}^{\rm event}\ d^2N_{\rm AA}/dy dp_T}
  {\left<N_{\rm coll}\right>/\sigma_{pp}^{\rm inelas}\ d^2
    \sigma_{pp}/dy dp_T}\ .
\end{equation}
for the most central (5\% centrality)
Au+Au collisions at RHIC.  The Cronin enhancement is visible at all
momenta.  Hadron formation time effects do play a substantial role in
the few GeV region, since heavier hadrons (K$^*$'s, $\rho$'s, protons)
are formed 7 times earlier than the rather light pions in the cms frame
at fixed transverse momentum due to the lower Lorentz boost $\gamma <5$.
It was shown in \cite{CGG} that for transverse momenta $p_T \geq 6$ GeV/c
the interactions of formed hadrons are not able to explain the
attenuation observed experimentally. However, the ratio $R_{\rm AA}$ is
influenced by interactions of formed (pre-)hadrons in the $p_T =
1\dots5\,{\rm GeV}/c$ range \cite{CGG}; a similar behaviour has also
been found in UrQMD simulations \cite{Bass_UrQMD}.

As pointed out before, the suppression seen in the calculation for
larger transverse momentum hadrons is due to the interactions of the
leading (pre-)hadrons with target/projectile nucleons and the bulk of
low momentum hadrons. It is clear that the experimentally observed
suppression can not be quantitatively described by the (pre-)hadronic
attenuation of the leading particles \cite{CGG}.  The ratio $R_{\rm
AA}$ (3) decreases to a value of about $0.5$ at 5 GeV for central
collisions, whereas the data are around $R_{\rm AA}\approx 0.25$.

For particles observable with momenta $p_T \geq 4$ GeV/c, the HSD
transport calculation predicts that still 1/3 of the final observed
hadrons have suffered one or more interactions, whereas the other 2/3
escape freely, i.e., without any interaction (even for central
collisions).  This implies that the final high $p_T$ hadrons originate
basically from the surface.

\subsection{Angular Correlations of Jets -- Can jets fake the large
$v_2$-values observed?}

\begin{figure}[h!]
\begin{minipage}[l]{7 cm}
\psfig{file=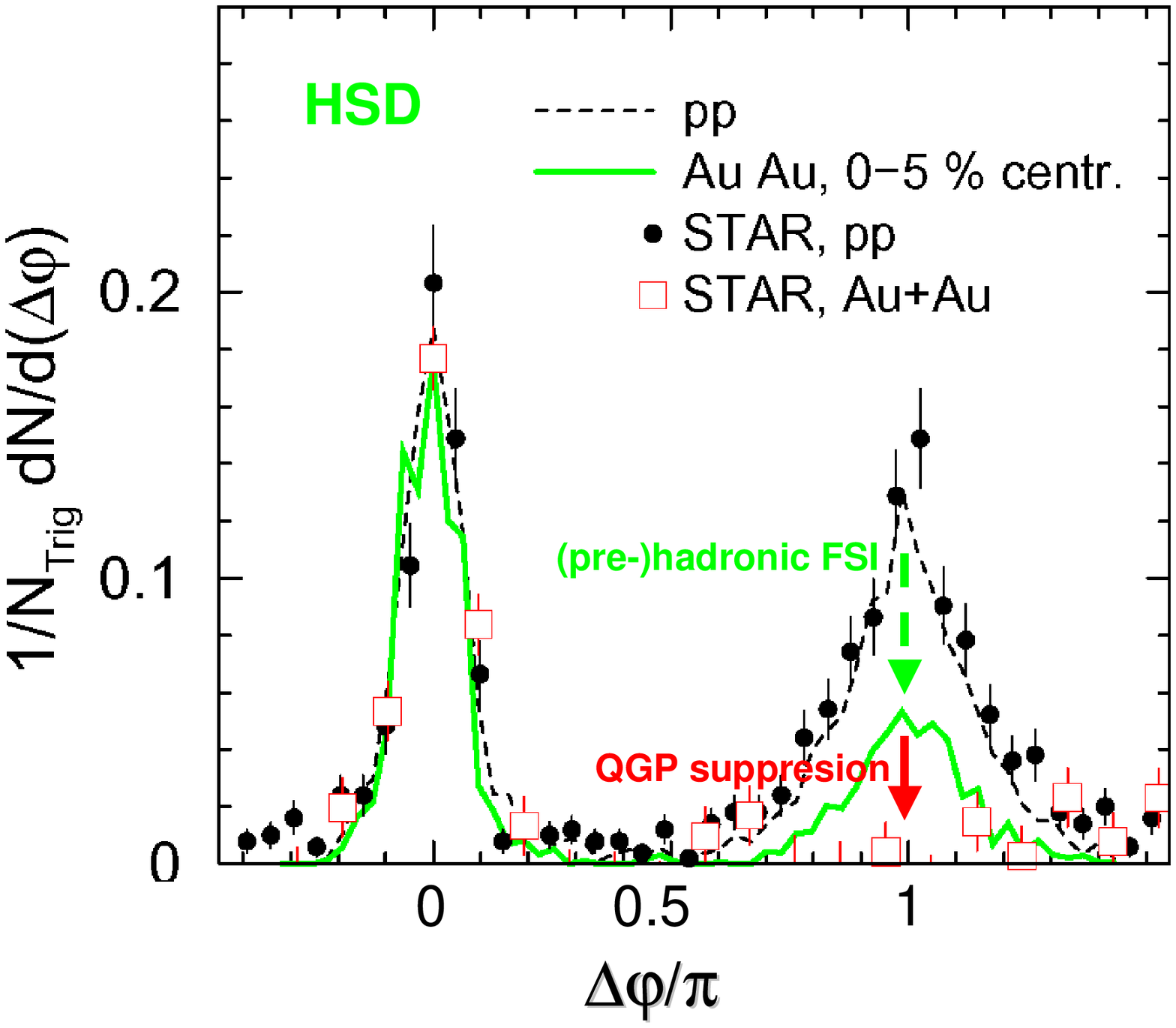,scale=0.43}
\end{minipage}
\begin{minipage}[r]{8 cm}
\vspace*{-2mm} \hspace*{1.2cm}
\epsfig{file=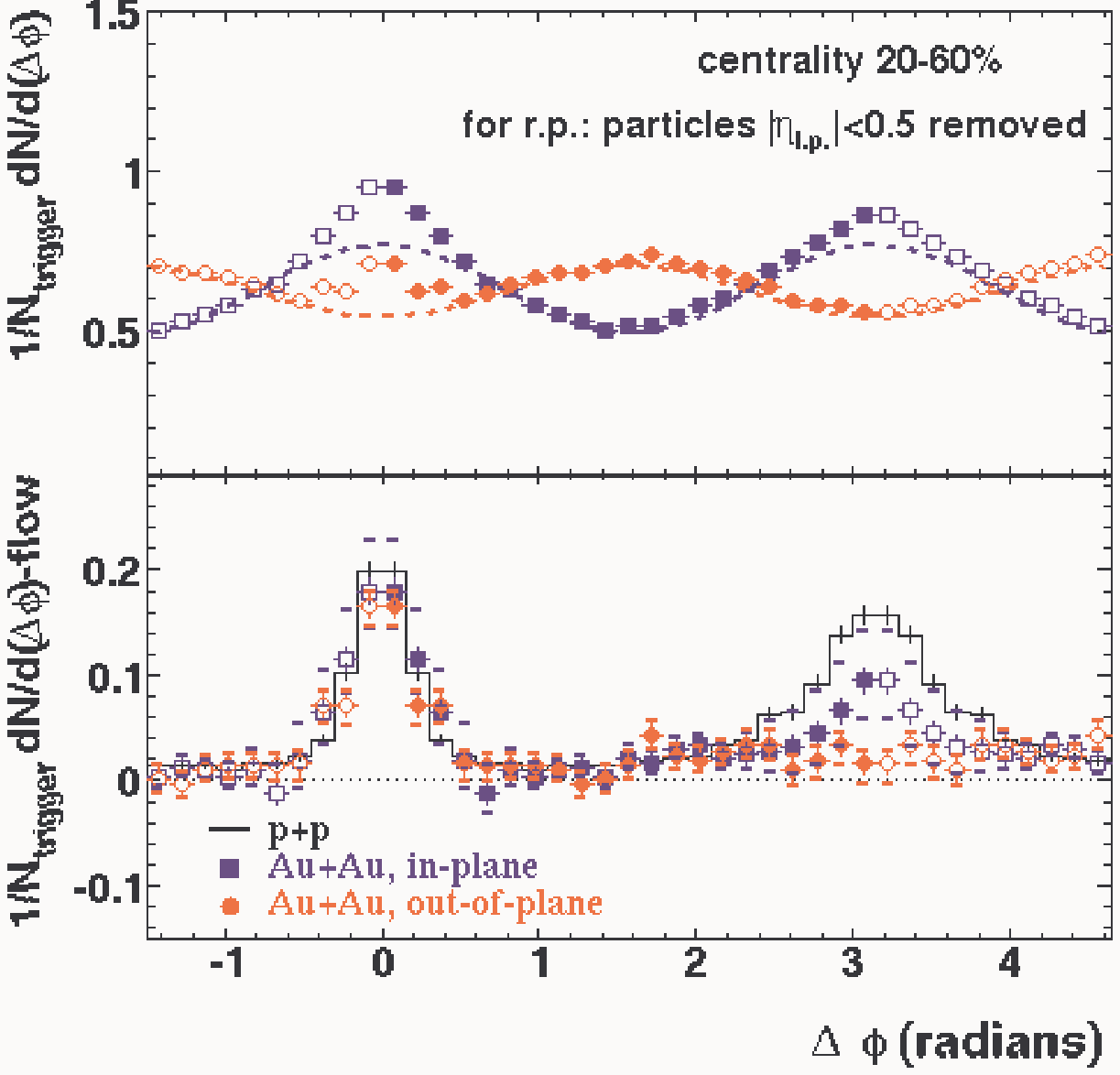,scale=0.53}
\end{minipage}
\caption{
Left: STAR data on near-side and away-side jet correlations compared
to the HSD model for p+p and central
Au+Au collisions at midrapidity for $p_T^{Trig}=4\dots6\,{\rm GeV}/c$ and
$p_T=2\,{\rm GeV}/c\dots p_T^{Trig}$ \protect{\cite{CGG,Cassing04}}.
Right: High $p_T$ correlations: in-plane vs. out-of-plane
correlations of the probe (jet+secondary jet fragments) with the bulk
($v_2$ of the plasma at $p_T > 2\,$GeV/c), prove the existence of the
initial plasma state (STAR-collaboration, preliminary).}
\label{angcorr}
\end{figure}

Fig.~\ref{angcorr} (l.h.s.) \cite{Cassing04} shows the angular
correlation of high $p_T$ particles ($p_T^{Trig}=4\dots6\,{\rm GeV}/c$,
\ \ $p_T=2\,{\rm GeV}\dots p_T^{Trig}$, $|y| <0.7$) for the 5\% most
central Au+Au collisions at $\sqrt{s}$ = 200 GeV (solid line) as
well as $pp$ reactions (dashed line) from the HSD model
\cite{Cassing04} in comparison to the data from STAR for $pp$
collisions \cite{StarAngCorr}.  Gating on high $p_T$ hadrons (in the
vacuum) yields 'near--side' correlations in Au+Au collisions close to
the 'near--side' correlations observed for jet fragmentation in the
vacuum (pp).  This is in agreement with the experimental observation
\cite{StarAngCorr}.  However, for the away-side jet correlations,  the
authors of Ref. \cite {Cassing04} get only a  $\sim$50\% reduction,
similar to HIJING, which has only parton quenching and neglects hadron
rescattering.  Clearly, the observed \cite{StarAngCorr} complete
disappearance of the away-side jet (Fig.~\ref{angcorr}) cannot be
explained in the HSD (pre-)hadronic cascade even with a small formation
time of $0.8\,$fm/c. Hence, the correlation data provide another  clear
proof for the existence of the bulk plasma.

Although (pre-)hadronic final state interactions yield a sizable ($\sim
50 \%$) contribution to the high $p_T$ suppression effects observed in
Au+Au collisions at RHIC, $\sim 50 \%$ of the jet suppression
originates from interactions in the plasma phase.  The elliptic flow,
$v_2$, for high transverse momentum particles is underestimated by at
least a factor of 3 in the HSD transport calculations \cite{CGG} (cf.
Fig.~\ref{v2_rhic}).  The experimentally observed proton excess over
pions at transverse momenta $p_T > 2.5$ GeV/c cannot be explained
within the CGG approach \cite{CGG}; in fact, the proton yield at high
$p_T \geq 5$ GeV/c is a factor 5-10 too small.  We point out that this
also holds for partonic jet-quenching models.

Futhermore, can the attenuation of jets of $p_T \ge5\,$GeV/c actually
fake the observed $v_2$-values at $p_T \approx 2\,$GeV/c? This question
comes about since due to fragmentation and rescattering a lot of
momentum-degraded hadrons will propagate in the hemisphere defined by
the jets. However, their momentum dispersion perpendicular to the jet
direction is  so large that it could indeed fake a collective flow that
is interpreted as coming from the high pressure early plasma phase (cf.
also Ref. \cite{Kovchegov}).

On first sight, Fig. \ref{angcorr} (r.h.s) shows that this could indeed  be
the case:  the in-plane $v_2$ correlations are aligned with the jet
axis, the away-side bump, usually  attributed to collective $v_2$ flow
(dashed line), could well be rather due to the stopped, fragmented and
rescattered away-side jet!  However, this argument is falsified by the
out-of-plane correlations (circles in r.h.s. Fig. \ref{angcorr}).  The
near-side jet is clearly visible in the valley of the collective flow
$v_2$ distribution.  Note that $v_2$ peaks atm $\varphi = \pi/2$
relative to the jet axis! The away-side jet, on the other hand, has
completely vanished in the out-of-plane distribution !

Where are all the jet fragments gone?  Why is there no trace left?
Even if the away-side jet fragments completely and the fragments get
stuck in the plasma, leftovers should be detected at momenta below
2 GeV/c.  Hadronic models as well as parton cascades will have a
hard time to get a quantitative agreement with these exciting data!

We propose future correlation measurements which can yield
spectroscopic information on the plasma:
\begin{enumerate}
\item
If the plasma is a colorelectric plasma, experiments will - in spite of
strong plasma damping - be able to search for wake-riding potential
effects. The wake of the leading jet particle can trap comoving
companions that move through the plasma in the wake pocket with the
same speed ($p_T/m$) as the leading particle. This can be particular
stable for charmed jets due to the deadcone effect as proposed by
Kharzeev et al \cite{Kharzeev}, which will guarantee little energy
loss, i.e. constant velocity of the leading D-meson. The leading
D-meson will practically have very little momentum degradation in the
plasma and therefore the wake potential following the D will be able to
capture the equal speed companion, which can be detected
\cite{Schafer78}.

\item
One may measure the sound velocity of the expanding plasma by the
emission pattern of the plasma particles travelling sideways with
respect to the jet axis: The dispersive wave generated by the wake of
the jet in the plasma yields preferential emission to an angle
(relative to the jet axis) which is given by the ratio of the leading
jet particles' velocity, devided by the sound velocity in the hot dense
plasma rest frame.  The speed of sound for a non-interacting gas of
relativistic massless plasma particles is $c_s \approx
\frac{1}{\sqrt{3}} \approx 57 \% \,c$, while for a plasma with strong
vector interactions, $c_s =  c$.  Hence, the emission angle measurement
can yield information of the interactions in the plasma.
\end{enumerate}

\section{Summary}

The NA49 collaboration has observed the collapse of both, $v_1$- and
$v_2$-collective flow of protons, in Pb+Pb collisions at 40 A$\cdot$GeV,
which presents first evidence for a first order phase transition in
baryon-rich dense matter. It should be possible to study the nature of
this transition and the properties of the expected chirally restored
and deconfined phase both at the forward fragmentation region at RHIC,
with upgraded and/or second generation detectors, and at the new GSI
facility FAIR.  According to Lattice QCD results
\cite{Fodor04,Karsch04}, the first order phase transition occurs for
chemical potentials above 400 GeV.  Thus, the observed collapse of
flow, as predicted in \cite{Hofmann74,Hofmann76}, is a clear signal for
a first order phase transition at the highest baryon densities.

A critical discussion of the use of collective flow as a barometer for
the equation of state (EoS) of hot dense matter at RHIC showed that
hadronic rescattering models can explain $< 30 \%$ of the observed
elliptic flow $v_2$ for $p_T > 2$ GeV/c.  We interpret this as
evidence for the production of superdense matter at RHIC with initial
pressure way above hadronic pressure, $p > 1$~GeV/fm$^3$.

The fluctuations in the flow, $v_1$ and $v_2$, should be measured.
Ideal Hydrodynamics predicts that they are larger than 50 \%  due to
initial state fluctuations.  The QGP coefficient of viscosity may be
determined experimentally from the fluctuations observed.

The connection of $v_2$ to jet suppression has been examined. It is proven
experimentally that the collective flow is not faked by minijet
fragmentation and theoretically that the away-side jet suppression can
only partially ($<$ 50\%) be due to pre-hadronic or hadronic
rescattering.

We propose upgrades and second generation experiments at RHIC, which
inspect the first order phase transition in the fragmentation region,
i.e. at $\mu_B\approx~400$~MeV  ($y \approx 4-5$), where the collapse
of the proton flow -- analogous to the 40 A$\cdot$GeV data --
should be seen.
Furthermore, the study of Jet-Wake-riding potentials and Bow shocks
caused by jets in the QGP formed at RHIC can give further clues on the
equation of state  and transport coefficients of the Quark Gluon
Plasma.
Moreover, we propose that the change in sign of $v_1, v_2$ closer to beam
rapidity is related to the occurence of a high density first order
phase transition in the RHIC data at 62.5, 130 and 200 A$\cdot$GeV.

\vspace*{3mm}
We like to thank W. Cassing, A. Dumitru, K. Gallmeister, C. Greiner,
K. Paech, A. Tang, N. Xu and Z. Xu for their contributions to this review.


\end{document}